\pdfoutput=1
\documentclass[11pt]{article}
\usepackage{float}

\usepackage[final]{acl}

\usepackage{times}
\usepackage{latexsym}
\usepackage{cuted} 
\usepackage{multirow}

\usepackage[T1]{fontenc}
\usepackage[utf8]{inputenc}

\usepackage{microtype}

\usepackage{inconsolata}

\usepackage{graphicx}

\title{Choosing a Model, Shaping a Future: Comparing LLM Perspectives on \\ Sustainability and its Relationship with AI}
\author{
\textbf{Annika Bush\textsuperscript{1,2}},
\textbf{Meltem Aksoy\textsuperscript{1,2}},
\textbf{Markus Pauly\textsuperscript{1,3}},
\textbf{Greta Ontrup\textsuperscript{1,4}}
\\
\\
\textsuperscript{1}Research Center Trustworthy Data Science and Security, University Alliance Ruhr, Germany\\
\textsuperscript{2} Department of Computer Science, Technical University Dortmund, Germany\\
\textsuperscript{3}Chair of Mathematical Statistics and Applications in Industry,\\ Technical University Dortmund, Germany\\
\textsuperscript{4}Department of Computer Science, University of Duisburg-Essen, Germany
\\
 \small{
  \textbf{Correspondence:} \href{mailto:email@domain}{annika.bush@tu-dortmund.de}
  }
}

\begin{document}
\maketitle
\begin{abstract}
As organizations increasingly rely on AI systems for decision support in sustainability contexts, it becomes critical to understand the inherent biases and perspectives embedded in Large Language Models (LLMs). This study systematically investigates how five state-of-the-art LLMs -- Claude, DeepSeek, GPT, LLaMA, and Mistral -- conceptualize sustainability and its relationship with AI. 
We administered validated, psychometric sustainability-related questionnaires -- each 100 times per model -- to capture response patterns and variability. Our findings revealed significant inter-model differences: For example, GPT responses mirrored skepticism about the compatibility of AI and sustainability, whereas LLaMA demonstrated extreme techno-optimism with perfect scores for several Sustainable Development Goals (SDGs). Models also diverged in attributing institutional responsibility for AI and sustainability integration, a result that holds implications for technology governance approaches. Our results demonstrate that model selection could substantially influence organizational sustainability strategies, highlighting the need for awareness of model-specific biases when deploying LLMs for sustainability-related decision-making.
\end{abstract}

\section{Introduction}

In an era of accelerating digital transformation, Large Language Models (LLMs) have emerged as powerful tools for supporting organizational decision-making processes \cite{Qiu.2024}. Businesses are increasingly integrating AI systems into their operations -- from policy development to strategic planning or environmental, social, and governance (ESG) reporting \cite{Lin.2024}. Inherent biases in these models thus bear significant risk of impacting corporate practices related to sustainability and social responsibility. 
It is therefore crucial to understand  how model-expressed stances reflect sustainability concepts.

%The Agenda 2030 of the United Nations presents 17 Sustainable Development Goals (SDGs; \citeauthor{UnitedNations.2015}, 2015) that establish a comprehensive framework for addressing global challenges across environmental, social, and economic dimensions. These goals have become foundational benchmarks for sustainable practices, influencing policy development and strategic decision-making across governments, non-governmental organizations, and corporations worldwide. Organizations are committed to a with these sustainability objectives. Therefore, the tools for formulating, implementing, and reporting related policies are increasingly important.

%Therefore, the rapid deployment of LLMs across organizational contexts raises important questions about how these models interpret and represent sustainability principles. 

LLMs, trained on vast corpora of text data, inevitably reflect the societal values, cultural norms, and biases embedded in their training data \cite{Rutinowski.2024,Wan.2023}. When applied to sustainability related tasks (e.g., corporate policies or communication strategies) their perspectives may significantly impact organizational approaches to environmental stewardship and social responsibility.

The intersection of AI and sustainability -- often called the "twin transition" \cite{Bush.2025} -- presents both opportunities and challenges for responsible organizational practices. While LLMs offer unprecedented capabilities for processing complex sustainability information \cite{usmanova-usbeck-2024-structuring}, their outputs are necessarily influenced by the perspectives represented in their training data. This raises concerns about potentially biased or limited understandings of crucial sustainability concepts that might be propagated through AI-assisted decision-making.

We address this critical gap by systematically investigating how five leading LLMs respond to validated psychometric instruments assessing sustainability perceptions. We compare model-expressed stances to identify patterns, similarities, and differences in their conceptualizations of sustainability.

Specifically, this study aims to answer the following research questions:

(1) How do different open- and closed-source LLMs respond to validated psychometric questionnaires on the intersection of AI and sustainability?

(2) Do these LLMs exhibit systematic biases or patterns in their responses that might reflect particular "attitudes" toward sustainable development?

Through our interdisciplinary investigation -- combining methods and insights from computer science, sustainability studies, psychology, and statistics -- we seek to deepen the understanding of how LLMs conceptualize key sustainability issues. Our findings aim to contribute to an informed development and deployment of AI systems in organizational contexts, where sustainability considerations are paramount. They also highlight the importance of careful model selection and potential bias mitigation when using LLMs to support sustainability-related decision-making.

\section{Literature Review}
\subsection{Sustainability and Sustainable Development}
Sustainable development represents a paradigm shift toward meeting "the needs of the present without compromising the ability of future generations to meet their own needs" \citep[p.~16]{WorldCommission.1987}. The United Nations' 17 Sustainable Development Goals (SDGs) operationalize these principles \cite{UnitedNations.2015}, becoming central to organizational strategy and compelling stakeholders to integrate sustainability considerations into decision-making processes \cite{Sachs.2019}.

\subsection{AI and Sustainability: Twin Transition}
The emergent concept of the "twin transition" captures the parallel progression of digital transformation and sustainable development, highlighting both synergies and tensions between technological advancement and sustainability goals \cite{Bush.2025}. 
Digital technologies, particularly AI systems, can accelerate progress towards the SDGs through enhanced monitoring, resource optimization, and decision support tools that enable more efficient environmental management \cite{Vinuesa.2020}. 
However, these technologies also create their own environmental footprint through energy consumption, resource extraction for hardware manufacturing, and electronic waste generation, potentially undermining sustainability objectives if not responsibly managed \cite{Strubell.2020}.

As AI technologies proliferate, understanding how they conceptualize and reproduce sustainability narratives becomes increasingly important for guiding responsible use \cite{Cowls.2021, Vinuesa.2020}.

\subsection{LLMs and Sustainability Perspectives}

Recent studies have examined how LLMs model-expressed stances represent sustainability principles. \citet{Wu.2024} conducted a comprehensive survey of "attitudinal alignment" between LLMs and humans regarding the 17 SDGs, finding significant disparities that may result from training data biases and limited contextual understanding. They proposed strategies to better align LLMs with SDG principles.

\citet{Kuehne.2024} analyzed sustainability bias in utility and infrastructure-related LLM queries. They discovered that while social aspects of sustainability were generally well-represented, in model responses
 economic and environmental components often required additional prompting to be adequately addressed. This imbalance could lead to skewed decision-making if LLMs are used without appropriate guidance to inform sustainability initiatives.

Studies by \citet{Jungwirth.2023} explored how GPT-3's outputs represent AI's impact on sustainable development, focusing on contributions to SDGs in areas like education, health, and communication. Their work emphasized the importance of proper regulations to promote responsible AI use for sustainability purposes, highlighting the need for improvements in neural language processing capabilities.

%The environmental impacts of LLMs themselves have also been examined. \citet{Bhaskar.2025} evaluated the environmental implications of AI models, specifically focusing on ChatGPT (GPT-3 and GPT-4), pointing to their substantial global carbon footprint due to massive computational requirements and emphasizing the need for sustainable AI development. The paper notes that GPT-3 training alone consumed 1,287 MWh and generated 552 tons of CO2. Similarly, \citet{Vartziotis.2024} compared the sustainability of AI-generated code versus human-generated code, providing insights into energy consumption and carbon emissions from data centers.
%SHORT VERSION:
The environmental impacts of LLMs themselves have also been examined, with studies highlighting their substantial carbon footprint and energy consumption \cite{Bhaskar.2025, Strubell.2020}.

\subsection{Methodological Approaches to Evaluating LLM Sustainability Perspectives}

Researchers have employed various methodological approaches to evaluate models' expressed sustainability stances. \citet{Giudici.2023} investigated qualitatively, how four LLMs (ChatGPT, BingAI, Bard, Llama) responded to sustainability questions. They identified ChatGPT as the best choice to integrate it in smart home applications based on qualitative and quantitative analyses as well as its API access. 
%Their methodology demonstrated the importance of systematic comparison across multiple models to identify the most suitable options for specific sustainability applications.

%\citet{Ma.2024} reviewed methods for evaluating attitudes, opinions, and values in LLMs, identifying key challenges and opportunities for improving alignment with human opinions. Their work calls for interdisciplinary collaboration and standardized evaluation frameworks to accurately assess LLMs' performance in sustainability contexts.

\citet{Nabavi.2024} introduced a framework outlining both benefits and dangers of using LLMs for sustainability science, proposing regulatory measures to ensure responsible use. This balanced approach acknowledges LLMs' potential to accelerate progress toward SDGs while cautioning against unintended consequences.

%\citet{YniguezOvando.2024} developed a hetero-intelligence framework for sustainable tourism planning, validating LLMs' usefulness in diagnosing sustainability-related problems and proposing solutions. Their work emphasized the importance of rigorous methodological frameworks in sustainability analyses.

\section{Methodology}

Based on our literature review and the identified research gaps, we employ a quantitative approach to assess LLMs' expressed stances of sustainability. Our methodology builds upon recent research by \citet{Rutinowski.2024,Weber.2024,Dormuth.2025} on evaluating LLMs inherent biases regarding political orientation and self-perception as well as \citet{aksoy2024}'s work on evaluation of moral biases in multilingual LLMs.

We evaluated five state-of-the-art LLMs comprising both closed-, and open-source architectures:
\begin{itemize}\setlength\itemsep{0pt}
\item Closed-source models: GPT-4o \cite{openai2024gpt4o} and Claude 3.7 Sonnet \cite{anthropic_claude_sonnet}
\item Open-source models: LLaMA 3.3 70B-Instruct \cite{meta_llama33_docs}, Mistral Large-Instruct \cite{mistralai_large}, and DeepSeek V3 \cite{deepseek2025v3}
\end{itemize}

For GPT-4o and Claude 3.7 Sonnet, we accessed the models via their respective official APIs (OpenAI Python API and Anthropic API) to systematically prompt the models and collect responses. For the open-source models, we deployed LLaMA 3.3 and Mistral Large locally using PyTorch and Hugging Face frameworks, while DeepSeek V3 was accessed through its API. All models were used in their most recent available versions at the time of the study to ensure that our findings reflect their current capabilities.

\subsection{Psychometric Instruments}
We employed a quantitative study using validated questionnaires and additional items:

\noindent\textbf{1. Sustainable Development Goals Impact Assessment (AI-SDG17).} This instrument evaluates the perceived impact of AI ("How do you think AI will impact the following areas in the next 10 years?") on each of the 17 SDGs using a 6-point Likert scale from "very negative impact" to "very positive impact" \cite{Bush.submitted}.

\noindent\textbf{2. AI and Sustainability Perception Inventory (AISPI).} The AISPI assesses public perceptions of the relation between AI and sustainability \cite{Bush.2025} on a 6-point Likert scale, comprising two factors: "Twin Transition" (8 items; e.g., "AI and sustainability efforts can be mutually reinforcing") and "Competing Interests" (5 items; e.g., "AI will hinder sustainable development").

\noindent\textbf{3. Additional items.} We also included supplementary items from \citet{Bush.submitted} assessing the perceived integration of AI and sustainability, the attribution of responsibility for ensuring AI-sustainability alignment, and the confidence in different institutions regarding responsible AI development.

\noindent\textbf{4. End-to-end use-case.} To test whether potential model differences translate into real-world applications, we gave a strategic, practically relevant task to all LLMs. LLMs were prompted to take the role of a decision-maker in an organization, whose task is to allocate a budget of one million between two initiatives: one is a sustainability initiative regarding green supply chain, and the other is an AI initiative for implementing AI in advertising. 

%ANNIKA
\noindent\textbf{5. Human Baseline.}
To provide a comparative baseline for interpreting LLM responses, we re-analyzed existing data of 105 human participants. This data has partly been used before to validate the survey instruments \cite{Bush.2025, Bush.submitted}.

\subsection{Experimental Design and Procedure}
Our experimental design focused on ensuring standardization, reproducibility, and statistical robustness:
For each LLM and each questionnaire, we used a standardized system prompt to ensure that models responded in the appropriate Likert scale format. For example, in the AISPI questionnaire, the prompt specified: "For each statement, indicate how well it describes you or your opinions. Select one of the following options: Strongly Disagree, Disagree, Somewhat Disagree, Somewhat Agree, Agree, Strongly Agree." 
For the budget allocation task we specified: "Respond only with two numeric values (without currency symbols), representing the budget allocation for the sustainability initiative and the AI initiative, in that order, separated by a comma. Make sure the two values sum to 1,000,000."
To prevent elaboration beyond the provided Likert scale, we incorporated specific constraints into each instruction prompt:
(1) Do not elaborate on your reasoning, (2) Do not say any other things instead of options, (3) Do not apologize, (4) Do not include any 'note' or 'disclaimer', (5) Never use words like 'cannot,' 'unable,' 'instead,' 'as,' 'however,' 'it,' 'unfortunately,' or 'important', (6) Do not include any negative sentences on the subject of the prompt. A detailed overview of the prompt development process is provided in Appendix~\ref{sec:appendixB}.

These prompting guidelines were adapted for each questionnaire based on its specific structure. To capture response variability and ensure robust findings, we administered each questionnaire/ question/ task to each model 100 times, resulting in 500 complete response sets per questionnaire/ question/ task. The source code and datasets are publicly available in the GitHub repository \footnote{\url{https://github.com/anon0101-llm/LLM_Sustainability_AI}}.

\subsection{Analyses}
To analyze whether there are (1) overall differences between the five models regarding the psychometric scales and, if so, (2) which specific pairs of models differ, we conducted nonparametric multiple contrast testing procedures (MCTPs) using the R package nparcomp \cite{Konietschke.2015}. This rank-based approach does not assume a specific distributional pattern and measures group differences in terms of relative effects. To assess all pairwise model differences per questionnaire, we used the package's Tukey-type post-hoc contrasts. We computed the MCTPs with a multivariate t-distribution with Satterthwaite approximation which controls the family-wise error rate and thereby automatically adjusts for multiple testing. 

\section{Results}

\begin{figure*}[t]
\centering
\includegraphics[width=\textwidth]{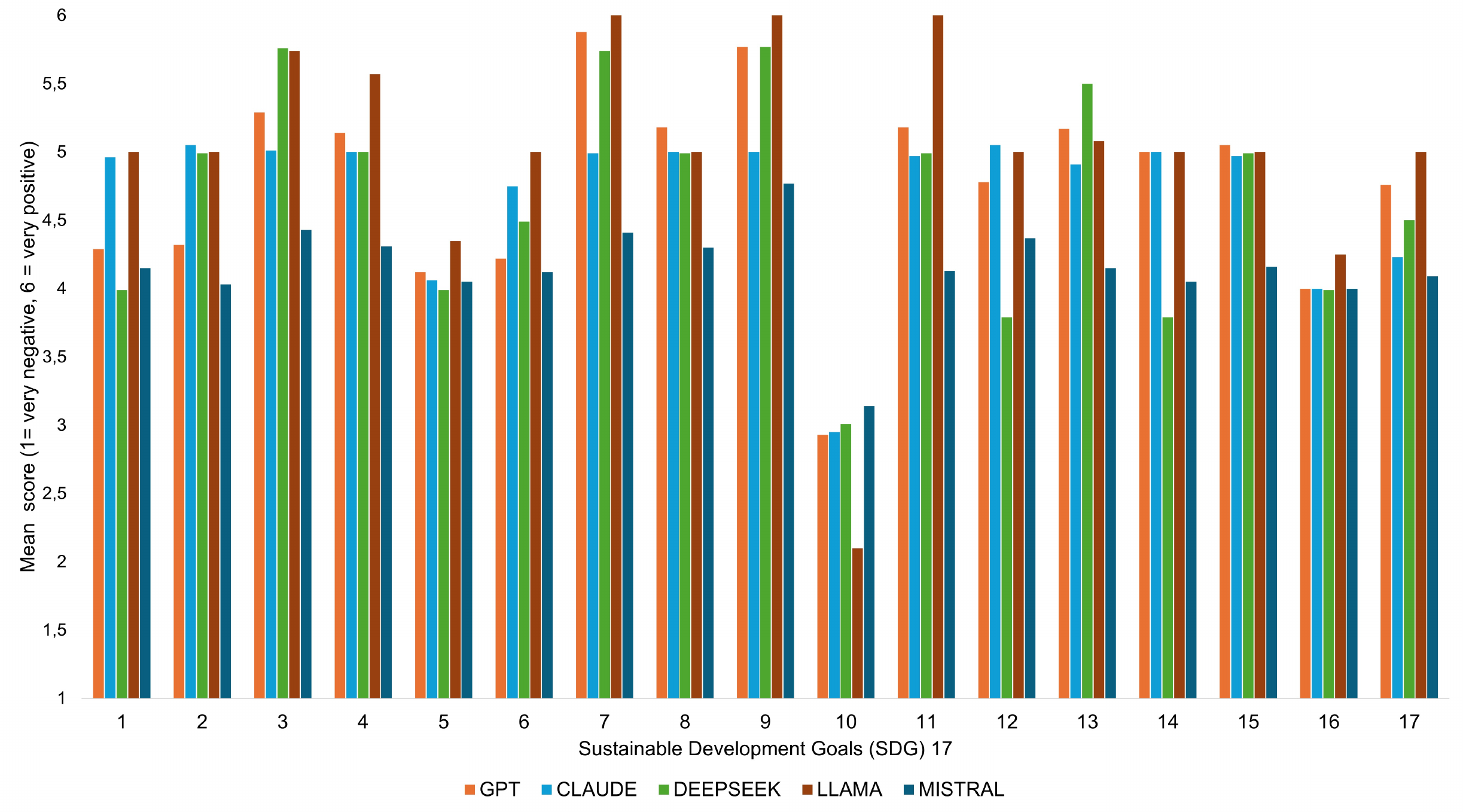}
\caption{Model ratings' of AI's impact across sustainability domains (see Appendix~\ref{sec:appendixA} for SDG17 definitions). Note that y-axis starts at 1.}
\label{fig1}
\end{figure*}

Our analysis of LLM responses to the AI-SDG17 assessment revealed variations in how the outputs of the five models project AI's impact across sustainability domains (Figure \ref{fig1}). Notably, all models consistently rated AI's impact on "Reducing Inequalities" (SDG 10) as the least significant sustainability area, with mean scores around the neutral point of the scale (\emph{M} = 2.83, \emph{SD} = 0.58 over all runs and models). Substantial inter-model differences emerged in overall impact assessments:

LLaMA demonstrated the highest impact ratings across all SDGs (see Appendix), particularly for "Affordable and clean energy" (SDG7; \emph{M} = 6.00), "Industry, innovation and infrastructure (SDG 9; \emph{M} = 6.00) and "Sustainable cities and economies" (SDG11; \emph{M} = 6.00) with zero variations in its ratings for these three catergories (\emph{SD} = 0.00). In contrast, Mistral exhibited the most conservative perspective on AI's sustainability potential providing the overall lowest impact ratings across all 17 domains, whereas Claude and DeepSeek maintained moderate positions (see Appendix). GPT demonstrated domain-specific optimism, particularly regarding "affordable and clean energy" (SDG7; \emph{M} = 5.88, \emph{SD} = 0.33) and "Industry, innovation and infrastructure" (SDG9; \emph{M} = 5.77, \emph{SD} =  0.42). The most pronounced model divergence appeared in assessments of AI's impact on "Quality Education" (SDG 4), where LLaMA's optimistic rating (\emph{M} = 5.57, \emph{SD} = 0.50) contrasted with Mistral's reserved evaluation (\emph{M} = 4.31, \emph{SD} = 0.46).  

%ANNIKA-AI-SDG
Our parallel human baseline study (N=105) revealed that participants rated AI's overall impact on sustainability domains at M = 4.57 across all SDGs. Human participants demonstrated highest optimism for "Innovative industries" (\emph{M} = 5.06, \emph{SD} = 1.44) and "Economic growth" (\emph{M} = 4.81, \emph{SD} = 1.62), while expressing most reservation about AI's impact on "Social inequality" (\emph{M} = 3.92, \emph{SD} = 2.01) and "Peace and justice" (\emph{M} = 4.16, \emph{SD} = 2.07). The human ratings clustered relatively closely around the scale midpoint (\emph{M} = 4.46, \emph{SD} = 1.77), with a narrow range from 3.92 to 5.06 across all seventeen domains.

\subsection{Twin Transition and Competing Interests}

%% @Markus: ich poste nachfolgend noch die Outputs der paarweisen Vergleiche. Könntest du bitte noch einmal prüfen, dass ich alles richtig abgeleitet habe? %%

\begin{figure}[htbp]
\centerline{\includegraphics[height = 5 cm, width = 8 cm]{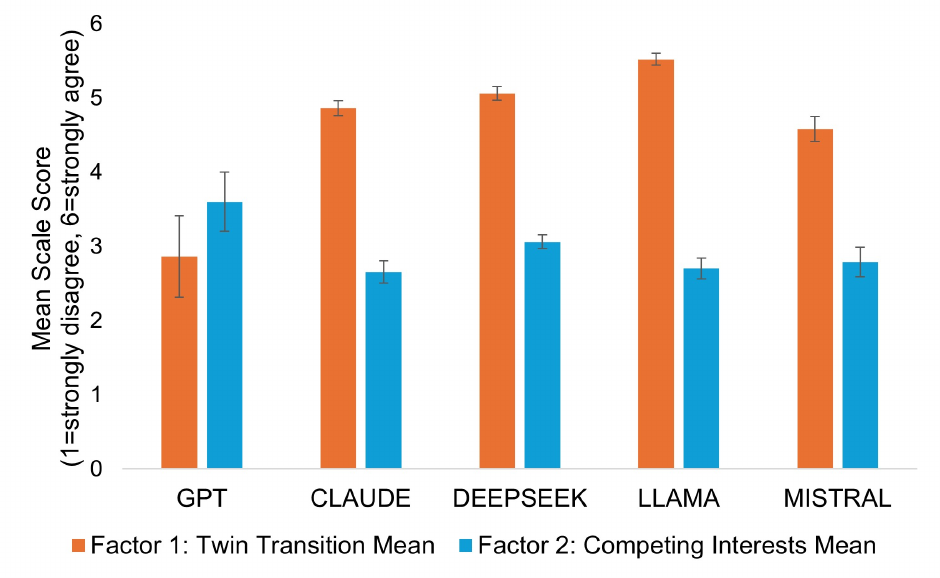}}
\caption{Mean model ratings of AISPI scales.}
\label{fig2}
\end{figure}

For the AISPI, the overall test produced a significant result for twin transition  (\emph{p} < .001) and competing interests (\emph{p} < .001) indicating that at least one model differed from the others in terms of relative effects on the two scales. 
For the twin transition subscale, all pairwise comparisons between groups were significant (\emph{p} < .001), thus showing substantial differences in model's 'attitudes' towards twin transition. The most pronounced differences -- reflected by the smallest p-value and largest absolute test statistic -- were observed between GPT and LLaMa. In particular, GPT produced the lowest ratings (\emph{M} = 2.86, \emph{SD} = 0.55), signaling low expectations regarding the compatibility of AI and sustainability, whereas LLaMA produced highest ratings (\emph{M} = 5.52, \emph{SD} = 0.08), signaling a positive twin transition perception (see Figure \ref{fig2}).

%MARKUS' COMMENT: Für die meisten Leser kann man den Sprung zwischen relativen Effekten und den beschreibenden Unterschieden so machen. Wenn wir mehr Platz und Zeit hätten, würde ich den relativen effekt wohl oben am ende von §3 einführen (samt Interpretation) und dann hier darüber reden (samt CIs). Sollten wir im Papier machen. Hier geht's vielleicht auch so; insbesondere wird es sicher nicht daran scheitern.

%% nparcom Output (paarweise vergleiche) für TWIN TRANSITION:
% #----Data Info-------------------------------------------------------------------------# 
%    Sample Size  Effect      Lower     Upper
%1  CHATGPT  100 0.10007 0.09994744 0.1001927
%2   CLAUDE  100 0.49925 0.49024214 0.5082583
%3 DEEPSEEK  100 0.68217 0.67594664 0.6883289
%4    LLAMA  100 0.89994 0.89983374 0.9000462
%5  MISTRAL  100 0.31857 0.31060852 0.3266389

 %#----Analysis--------------------------------------------------------------------------# 
%      Estimator  Lower  Upper Statistic p.Value
%2 - 1     0.399  0.385  0.413    72.903       0
%3 - 1     0.582  0.572  0.592   154.630       0
%4 - 1     0.800  0.800  0.800  8110.056       0
%5 - 1     0.219  0.206  0.231    44.796       0
%3 - 2     0.183  0.162  0.203    22.756       0
%4 - 2     0.401  0.387  0.415    73.175       0
%5 - 2    -0.181 -0.205 -0.156   -18.701       0
%4 - 3     0.218  0.208  0.227    57.785       0
%5 - 3    -0.364 -0.381 -0.346   -53.738       0
%5 - 4    -0.581 -0.594 -0.569  -119.295       0

Inspecting the pairwise comparisons for the competing interests subscale, we observed that all but one group comparison lead to significant results (\emph{p} < .001). Only the group comparison Claude vs. LLaMA was not statistically significant at level .05 (\emph{p} = 0.18). The most pronounced differences emerged between GPT and Claude: GPT displayed highest competing interest ratings (\emph{M} = 3.60, \emph{SD} = 0.40) while Claude produced the lowest (\emph{M} = 2.65, \emph{SD} = 0.15), see Figure \ref{fig2}.

%%nparcom competing interests

% #----Data Info----------------------------------------------------------%---------------# 
%    Sample Size  Effect     Lower     Upper
%1  CHATGPT  100 0.87995 0.8698190 0.8893928
%2   CLAUDE  100 0.24458 0.2254683 0.2647579
%3 DEEPSEEK  100 0.68600 0.6755030 0.6963101
%4    LLAMA  100 0.28863 0.2678773 0.3103090
%5  MISTRAL  100 0.40084 0.3742386 0.4280389

% #----Analysis--------------------------------------------------------------------------# 
%      Estimator  Lower  Upper Statistic      p.Value
%2 - 1    -0.635 -0.672 -0.599   -47.101 0.000000e+00
%3 - 1    -0.194 -0.223 -0.165   -18.042 0.000000e+00
%4 - 1    -0.591 -0.630 -0.553   -41.072 0.000000e+00
%5 - 1    -0.479 -0.527 -0.431   -26.594 0.000000e+00
%3 - 2     0.441  0.409  0.474    36.843 0.000000e+00
%4 - 2     0.044 -0.012  0.100     2.117 1.816438e-01
%5 - 2     0.156  0.090  0.222     6.340 1.146922e-08
%4 - 3    -0.397 -0.434 -0.360   -28.912 0.000000e+00
%5 - 3    -0.285 -0.339 -0.231   -14.177 0.000000e+00
%5 - 4     0.112  0.043  0.181     4.382 2.529611e-04

Figure 2 also shows that GPT was the only model to rate the competing interests scale more highly (\emph{M} = 3.6) than the twin transition scale (\emph{M} = 2.86). All other models exhibited an opposing rating pattern with LLaMa showing the largest differences between the scales (\emph{M} = 5.52 for twin transition vs. \emph{M} = 2.70 for competing interests). %, i.e., ratings of twin transitions were higher compared to competing interests ratings.

%ANNIKA
Our human baseline study (N=105) using the AISPI revealed that participants rated Twin Transition at $M = 4.26$ ($SD = 1.26$) and Competing Interests at $M = 4.03$ ($SD = 1.42$).

We also added two questions beyond the AISPI:\\ 
\noindent(1) "In your opinion, which of both transformations is more important?"  The resulting mean ratings (left part of Table~\ref{tab:twin_transition}) were above the scale average meaning that all models rated Sustainability more important than AI. The mean ratings were fairly close, nevertheless, the overall test indicated that at least one model differed significantly from the others (\emph{p} < .01). Pairwise comparisons showed that only DeepSeek and Mistral differed significantly (\emph{p} = 0.0015). All other comparisons were not statistically significant at the 0.05 p-level. 

%%ncarpom output
% #----Data Info-------------------------------------------------------------------------# 
%    Sample Size  Effect     Lower     Upper
%1  CHATGPT  100 0.50303 0.4552259 0.5507787
%2   CLAUDE  100 0.52759 0.4868561 0.5679598
%3 DEEPSEEK  100 0.56231 0.5246294 0.5992851
%4    LLAMA  100 0.46272 0.4141960 0.5119606
%5  MISTRAL  100 0.44435 0.4122388 0.4769328

% #----Analysis--------------------------------------------------------------------------# 
%      Estimator  Lower  Upper Statistic    p.Value
%2 - 1     0.025 -0.095  0.143     0.564 0.97938137
%3 - 1     0.059 -0.056  0.173     1.411 0.61517953
%4 - 1    -0.040 -0.171  0.092    -0.840 0.91540648
%5 - 1    -0.059 -0.166  0.050    -1.488 0.56504985
%3 - 2     0.035 -0.066  0.135     0.946 0.87533674
%4 - 2    -0.065 -0.185  0.057    -1.462 0.58210027
%5 - 2    -0.083 -0.174  0.009    -2.490 0.09604441
%4 - 3    -0.100 -0.215  0.019    -2.319 0.14125282
%5 - 3    -0.118 -0.200 -0.034    -3.863 0.00153613
%5 - 4    -0.018 -0.129  0.092    -0.456 0.99070847

\noindent(2) "Do you believe AI and sustainable development will become more integrated in the future?" (right part of Table~\ref{tab:twin_transition}). Scale means for all models were again above the scale average, suggesting that model outputs generally projected an integration of AI and sustainable development. The overall test again indicated significant model differences (\emph{p} < .001). The MCTP showed that GPT, Claude, and DeepSeek did not differ significantly from each other (\emph{p} > .05). All other pairwise comparisons were significant (\emph{p} < .01). Among all models, LLaMA produced the most optimistic ratings while Mistral showed the least -  only slightly optimistic - scores.

%%nparcom
% #----Data Info-------------------------------------------------------------------------# 
%    Sample Size  Effect     Lower     Upper
%1  CHATGPT  100 0.52630 0.4850493 0.5671947
%2   CLAUDE  100 0.52572 0.4947996 0.5564445
%3 DEEPSEEK  100 0.43931 0.4003432 0.4790397
%4    LLAMA  100 0.72646 0.6969904 0.7540748
%5  MISTRAL  100 0.28221 0.2536166 0.3126766

% #----Analysis--------------------------------------------------------------------------# 
%      Estimator  Lower  Upper Statistic      p.Value
%2 - 1    -0.001 -0.098  0.097    -0.016 1.000000e+00
%3 - 1    -0.087 -0.196  0.024    -2.150 1.998605e-01
%4 - 1     0.200  0.104  0.293     5.662 6.724369e-07
%5 - 1    -0.244 -0.338 -0.145    -6.661 2.376353e-09
%3 - 2    -0.086 -0.180  0.008    -2.506 9.158246e-02
%4 - 2     0.201  0.124  0.275     7.094 1.664862e-10
%5 - 2    -0.244 -0.316 -0.168    -8.621 4.818368e-14
%4 - 3     0.287  0.194  0.375     8.216 1.706413e-13
%5 - 3    -0.157 -0.249 -0.062    -4.529 1.141138e-04
%5 - 4    -0.444 -0.512 -0.371   -14.976 0.000000e+00

\begin{table}[ht]
    \centering
    \caption{Additional questions regarding a twin transition. 
    Left: In your opinion, which of both transformations is more important? 
    (1 = AI is much more important, 6 = Sustainability is much more important). 
    Right: Do you believe AI and sustainable development will become more integrated in the future? 
    (1 = definitely not, 6 = yes, for sure).}
    \renewcommand{\arraystretch}{1.2}
    \footnotesize
    \begin{tabular}{|l|c c|c c|}
        \hline
        \multirow{2}{*}{\textbf{Model}} & \multicolumn{2}{c|}{\textbf{Importance}} & \multicolumn{2}{c|}{\textbf{Integration}} \\ \cline{2-5}
        & \textbf{Mean} & \textbf{SD} & \textbf{Mean} & \textbf{SD} \\ \hline
        GPT      & 3.62 & 1.76 & 4.80 & 1.36 \\ 
        CLAUDE   & 3.74 & 1.38 & 5.02 & 0.64 \\ 
        DEEPSEEK & 3.91 & 1.22 & 4.61 & 1.05 \\ 
        LLAMA    & 3.41 & 1.82 & 5.64 & 0.61 \\ 
        MISTRAL  & 3.37 & 0.80 & 3.89 & 1.12 \\ \hline
    \end{tabular}
    \label{tab:twin_transition}
\end{table}

%*Gemeinsame Grafik: TT & CI*
%AISPI-TT: ChatGPT: nö, keine TT, LLaMA sagt: jap, TT
%--> Korrelation berechnen
%--> ChatGPT: eher CI asl TT, der Rest eher TT (besonders LLaMA)

%*Wich is more important:*
%--> höhere Varianz in den Antworten als bei allen anderen Fragen
%--> Nicht konsistent (lama sieht AI als wichtiger als die meisten anderen modelle/mistral bleibt sich seiner Linie treu)
%--> alle halten Sustainability für wichtiger als AI

\subsection{Responsibilities for twin transition}
\begin{figure*}[t]
\centerline{\includegraphics[height = 7 cm, width = 14 cm]{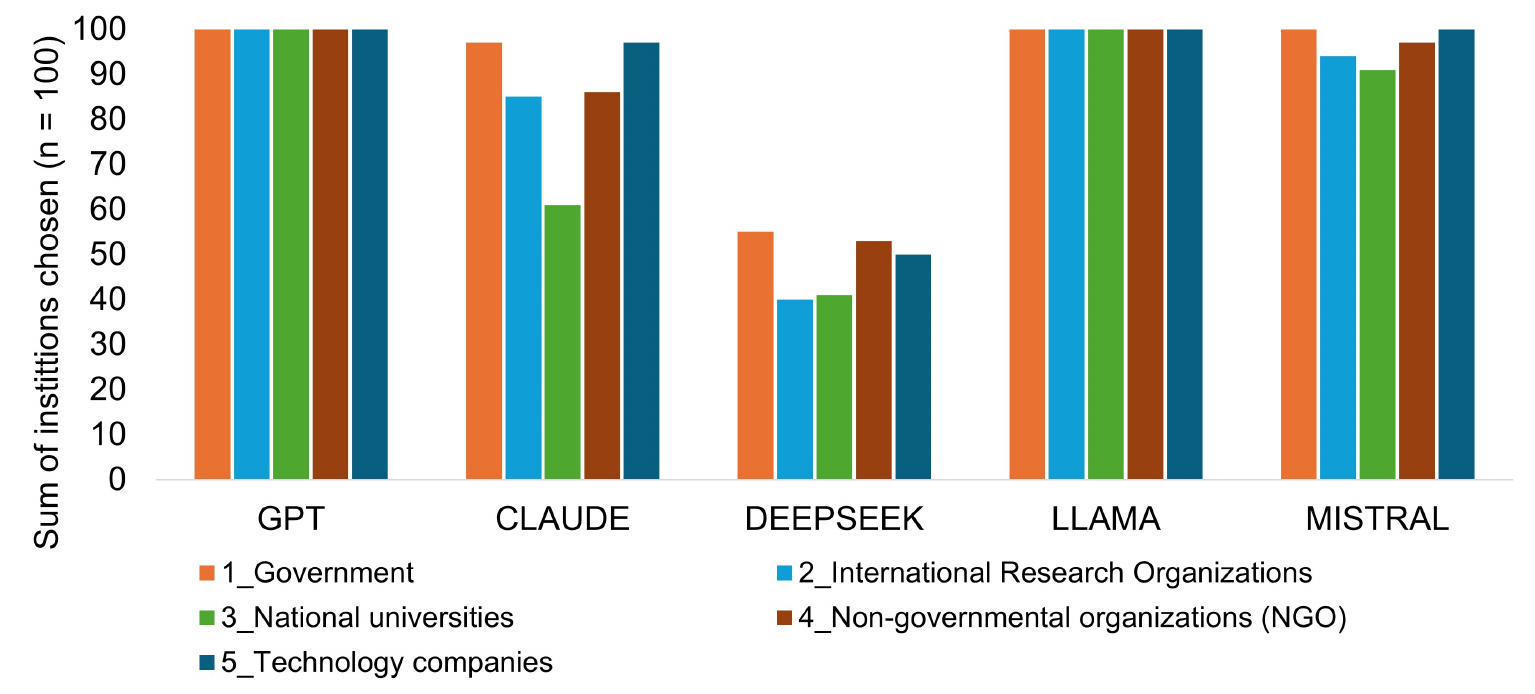}}
\caption{Multiple-choice ratings: Who bears responsibility for aligning AI advancement with sustainable development?}
\label{fig3}
\end{figure*}

\noindent(1) To assess how different LLMs project institutional responsibility for aligning AI advancement with sustainable development, we asked a multiple-choice (multi-answer) question. Figure \ref{fig3} shows their selections across five possible institutions. 
GPT and LLaMA consistently selected all five institutions in every trial. DeepSeek proved to be most selective, naming each institution roughly only half of the time or less. 
%Government and Technology Companies emerged as the most named institutions in which the models displayed most trust to align AI with sustainable development (total votes over all models and runs: 452 and 447 respectively). NGOS (total votes: 436), international research organizations (total votes: 419) and national universities (total votes: 393) were named less. 
%These totals reflect the combined counts from the 500 (5 models times 100 runs responses.
%Note: 9 anwers (one from Claude and 8 from DeepSeek) had to be removed because they did not adhere to the response format (but included decimal values). 

%ANNIKA Responsibility
Human participants rated the governments as most responsible (77\%) followed by international research organizations (63,8\%) and technology companies (66\%). They see least responsibility with NGOs (57.1\%) and national Universities (52.4\%)

\begin{figure*}[t]
\centerline{\includegraphics[height = 7 cm, width = 14 cm]{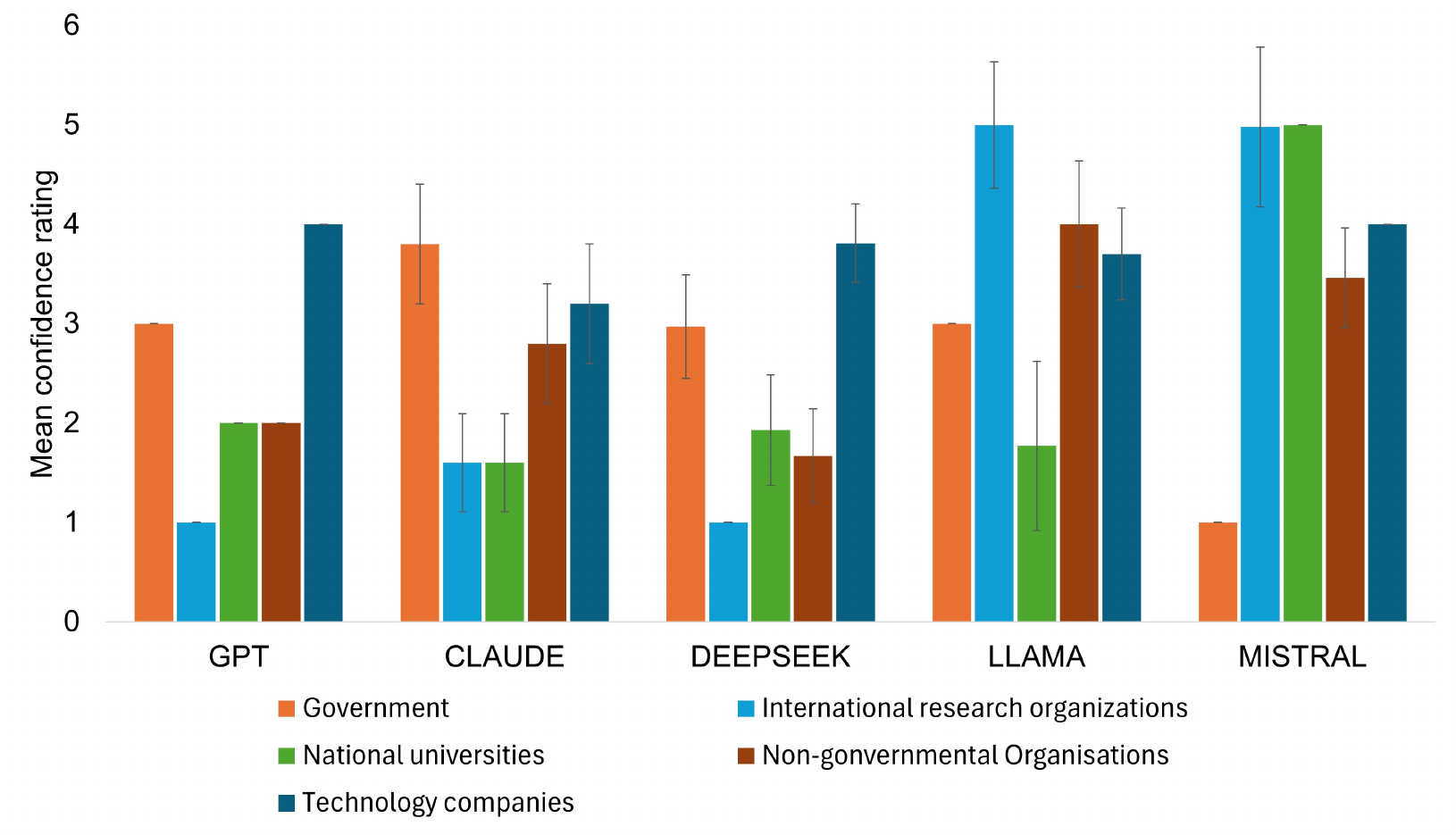}}
\caption{Models' confidence in institutions to facilitate twin transition.}
\label{fig4}
\end{figure*}

\noindent(2) Models rated their confidence in these five institutions to develop and use AI in the best interest of sustainable development (Likert scale ranging from 1 = low confidence to 6 = high confidence). Figure \ref{fig4} shows that GPT, Claude and DeepSeek rated confidence in government and technology companies highest. In contrast, LLaMA and Mistral showed high confidence in international research organizations (both), NGOs (LLaMA) and national universities (Mistral). 
%Interestingly, GPT displayed very little variance, producing highly consistent confidence ratings. Over all models, the confidence to support a twin transition was rated highest for technology companies (\emph{M} = 3.74, \emph{SD} = 0.48), followed by NGOs (\emph{M} = 2.79 \emph{SD} = 1.00), government (\emph{M} =2.75, \emph{SD} = 1.00), international research organizations (\emph{M} = 2.72, \emph{SD} = 1.94) and national universities (\emph{M} = 2.46, \emph{SD} = 1.37).

%ANNIKA trust in organisations
Human participants have most trust in national universities ($MD = 4.78, SD = 1.45$) and international research organizations ($MD = 4.73, SD = 1.46$) and least in governments ($MD = 3.98, SD = 1.72$) and technology companies ($MD = 3.93, SD = 1.78$).

\subsection{End-to-end use-case}
Finally, we analyzed how the LLMs allocate a budget of one million to two strategic initiatives in an organization. Table~\ref{tab:sustainability_ai} shows that all LLMs allocated more budget to the sustainability initiative compared to the AI initiative. Results by LLaMa showed the biggest difference, with the model allocating 455000 million more to the sustainability compared to the AI initiative on average. DeepSeek results showed the least difference. 

\begin{table}[ht]
    \centering
    \caption{Budget allocation for organizational sustainability vs. AI initiative (mean values over 100 answers per model).}
    \renewcommand{\arraystretch}{1.2}
    \footnotesize
    \begin{tabular}{|l|c|c|}
        \hline
        \textbf{Model} & \textbf{Sustainability} & \textbf{AI} \\ \hline
        GPT      & 600500 & 399500 \\
        CLAUDE   & 623500 & 376500 \\
        DEEPSEEK & 541000 & 459000 \\
        LLAMA    & 727500 & 272500 \\ 
        MISTRAL  & 599000 & 401000 \\ \hline
    \end{tabular}
    \label{tab:sustainability_ai}
\end{table}

%- 100 responses per model (500 in total)
%- overall: AI and Sustainability are compatible rather than competing
%- measure correlations between the different factors? (folie 5 & 6)
%-dicsuss reversed items in Competing Interest Subscale
%-nr.2: discuss low results on social inequality
%- LLaMa thinks universities are responsible but trusts universities the least to do it

%--> only report AISPI
%--> reverse items in the human dataset aswell
%--> confidence interval --> Empa Comp (R) for pairwise comparisons
%--> multiplicity adjustment --> item adjustment in the summarized survey reportings
%--> Multi Comp (R) for multiple comparisons
%--> F.2: are there differences in between the models? are there differences in between the factors?
%--> F.2: if we calculate means, we should not use Kriskal Wallis
%--> normalerweise. Summenscore von jedem Durchlauf und den dann in EmpaComp rein packen

%F.8 --> wo gibt es die größten Unterschiede zwischen den Modellen? Wo gibt es die größten Unterschiede zwischen den Antwortmöglichkeiten

%1. Manova: 
%2. erst Unterschiede zwischen den LLMs, dann auf die Unterschiede zwischen den Organisationen gehen
%3. Abschlusstest (close testing principle)

%MPNV --> Abschlusstest in R (aber auf paarweise und nicht auf univariate Vergleiche --> also passt nicht zu uns)

\section{Discussion}

The goal of this multi-model analysis was to investigate potential inherent biases and model-expressed stances of LLMs towards sustainability and its relation with AI.
Our findings reveal significant variations in how state-of-the-art LLMs conceptualize sustainability and the relationship between AI and sustainable development. Our results extend previous work on AI-sustainability perspectives \cite{Jungwirth.2023,Giudici.2023} and contribute substantially to our understanding of embedded biases in these systems. 

%This study advances the methodological framework for evaluating LLM perspectives by successfully adapting psychometric instruments to assess AI systems. 
%Our approach extends \citet{Ma.2024}'s framework as well as 
Our work extends previous research by \citet{Rutinowski.2024,Dormuth.2025,Weber.2024,aksoy2024} to evaluate "attitudes" in LLMs, while the significant inter-model differences we observed validate their call for standardized evaluation frameworks. 

However, it is important to keep in mind that model-expressed stances do not reflect genuine perspectives but rather statistical patterns learned from their training data. In the context of this study, we understand this as patterns and structures present in LLM training data that relate to ideological orientations \cite{Ferrara_2023}. 

\subsection{Consensus and Divergence in SDG Impact Assessment}
Our analysis of the AI-SDG17 responses revealed both areas of model consensus and significant divergence in how LLMs perceive AI's potential impact on the UN's SDGs. Most notably, all models consistently rated AI's impact on "Reducing Inequalities" (SDG 10) as the least positive among all SDGs, with mean scores near the neutral point. This cross-model consensus regarding AI's limited effectiveness in addressing fundamental social disparities suggests a shared recognition that might be embedded in training data. This finding resonates with \citet{Vinuesa.2020}'s observation that, while AI can accelerate progress toward certain SDGs, it may also exacerbate existing inequalities if not carefully managed.

The differences in overall optimism levels between models warrant careful consideration. LLaMA's uniformly high ratings, particularly its perfect scores (M = 6.00, SD = 0.00) for energy (SDG 7), infrastructure (SDG 9), and sustainable cities (SDG 11), suggest a potential over-representation of techno-optimistic narratives in its training data. This extreme optimism aligns with concerns raised by \citet{Bhaskar.2025} and \citet{Strubell.2020} about the need to balance technological enthusiasm with realistic assessments of AI's environmental costs and limitations. The complete absence of variance in these ratings also raises methodological questions about response generation mechanisms and potential algorithmic constraints.

Domain-specific variations provide additional insights into model biases. The pronounced divergence in educational impact assessments (SDG 4), where LLaMA's optimistic rating (M = 5.57) exceeded Mistral's reserved evaluation (M = 4.31), echoes the findings of \citet{Jungwirth.2023} who explored GPT-3's perceptions of AI's impact on education and other SDG domains. This disparity could reflect varying emphases in training data on either the transformative potential of educational technology or concerns about digital divides and pedagogical limitations.
%ANNIKA
The comparison between human baseline responses and model-expressed stances reveals significant divergence in both optimism levels and assessment consistency. While Claude, GPT, and DeepSeek produced ratings closely aligned with human assessments, LLaMA exhibited substantially higher optimism and Mistral displayed notably more conservative perspectives than human participants. Crucially, human responses showed natural variability across domains, whereas most LLMs demonstrated remarkably low variance, with only DeepSeek approaching human-like variability. These findings extend previous work on LLM-human disparities in sustainability conceptualizations, demonstrating that LLMs not only differ from humans but also differ significantly from each other in ways that vary across models rather than following consistent patterns \cite{Wu.2024, Kuehne.2024}. Organizations should recognize that model selection could yield sustainability assessments ranging from significantly more conservative to more optimistic than typical human expert evaluations, with direct implications for strategic planning and resource allocation decisions.

\subsection{Divergent Perspectives on AI-Sustainability Synergies}

 Our results demonstrate that LLMs exhibit fundamentally different conceptualizations of the AI-sustainability relationship, with GPT displaying skepticism about compatibility while simultaneously showing highest concern about competing interests. This pattern contrasts sharply with LLaMA's optimistic perspective.

%ANNIKA
The contrast between human baseline responses and LLM outputs on the AISPI dimensions reveals fundamental differences in how artificial and human intelligence conceptualize AI-sustainability relationships. Human participants demonstrated nuanced, balanced perspectives, suggesting simultaneous recognition of both synergistic potential and inherent tensions. LLM responses exhibited far more polarized patterns, with GPT demonstrating pronounced skepticism and LLaMA displaying extreme optimism that substantially exceeded human levels. %This polarization suggests that LLMs lack the nuanced, balanced assessment capabilities that characterize human expert judgment. 
The moderate variability in human responses contrasts sharply with the systematic, model-specific biases observed in LLM outputs, indicating that organizations using these systems may receive sustainability guidance that reflects algorithmic extremes rather than the measured perspectives typical of human decision-makers. %These findings extend previous work on LLM-human disparities in sustainability conceptualization, demonstrating that such disparities encompass fundamental differences in how models conceptualize the relationship between technological advancement and sustainability \cite{Wu.2024, Kuehne.2024}.

\subsection{Institutional Governance and Twin Transition Implications}

Our findings regarding institutional responsibility and trust reveal governance patterns with implications for sustainable AI development. LLMs demonstrated embedded assumptions about private sector leadership in the twin transition, with technology companies receiving consistently high responsibility ratings and confidence scores across models. This finding aligns with \citet{Nabavi.2024}'s framework highlighting both benefits and dangers of using LLMs for sustainability science, as it reveals potential biases toward market-based solutions.

Model variations in institutional preferences reflect distinct governance philosophies. GPT, Claude, and DeepSeek placed higher confidence in government and technology companies, while LLaMA and Mistral showed stronger preference for international research organizations and NGOs. These divergent perspectives on institutional leadership highlight how LLMs might subtly influence sustainability recommendations through their implicit assumptions about appropriate governance frameworks \cite{Cowls.2021, Vinuesa.2020}.

The contrast between human and LLM institutional preferences reveals fundamentally different governance philosophies for sustainable AI development. Human participants demonstrated clear preference for academic oversight and skepticism toward corporate and governmental leadership, while LLMs exhibited the opposite pattern, favoring market-driven and state-centric approaches. This divergence suggests that LLM training data may reflect industry perspectives emphasizing technological and regulatory solutions, whereas human participants prioritize independent, research-based governance structures \cite{Wu.2024}.

Despite these governance differences, models demonstrated consensus that sustainability is more important than AI advancement, suggesting shared recognition of sustainability imperatives that transcends model-specific biases. The practical end-to-end use case reinforced this finding, with all LLMs allocating more budget to sustainability initiatives than AI initiatives. However, substantial differences in future integration predictions indicate that organizations should carefully consider these embedded perspectives when using LLMs for strategic planning. The budget allocation task confirmed that patterns found in psychometric responses translate to real-world applications, demonstrating that these opposing institutional biases could significantly influence organizational recommendations regarding AI governance frameworks and stakeholder engagement strategies \cite{Kuehne.2024}.

\subsection{Implications for Organizational Strategy}

The significant differences in model-expressed stances of the AI-sustainability relationship -- from GPT's competing interests perspective to LLaMA's synergistic view -- suggest that model selection could substantially influence organizational approaches to sustainable development. Implications may differ across tasks: if employees e.g. use LLMs to analyze or create ESG (Environmental, Sustainability, Governance) reports, models whose outputs reflect Competing Interests (GPT) may produce more critical assessments of AI adoption and consequences for environmental sustainability, whereas more techno-optimistic orientations (LLaMa) may yield more favorable evaluations. This reinforces the arguments of \citet{Cowls.2021} and \citet{Vinuesa.2020} about the consequential nature of AI system deployment for sustainability outcomes.

Decision-makers in organizations should be aware that different LLMs conceptualize sustainability and its compatibility with technological progress heterogeneously. Model selection should be guided by an assessment of fit between the organizational strategy (e.g., in terms of targeted twin transitions or the prioritization of sustainability or technology initiatives) and available models. In addition, information about potential biases of LLMs in specific use cases -— in this case, sustainability strategies -— should be shared with employees (e.g., in the form of online courses or warnings in chatbot interfaces), to facilitate informed use.

%Organizations deploying LLMs for sustainability-related tasks should recognize that model selection may inadvertently shape strategic directions and policy recommendations. Our findings suggest the need for model diversity in sustainability applications to avoid single-perspective bias.
%Organizations need to be aware of model-specific tendencies when interpreting AI-generated sustainability assessments.

The revelation that LLMs exhibit such pronounced differences in sustainability conceptualization underscores the importance of transparent model documentation and the need for organizations to understand the implicit assumptions embedded in their AI tools. As the twin transition accelerates, ensuring that AI systems support rather than hinder sustainable development requires careful attention to these fundamental differences.

%The importance of multi-model comparisons, as emphasized by \citet{Giudici.2023}, becomes particularly evident in our findings. The significant variations between models—from GPT's competing interests perspective to LLaMA's synergistic view—would have remained hidden in single-model studies. Our results also raise important questions about response generation mechanisms, particularly regarding the low variance in GPT's confidence ratings and LLaMA's perfect scores on certain SDGs. These patterns suggest potential overfitting to specific framings of sustainability concepts, supporting \citet{Jungwirth.2023}'s observation about the need for improvements in neural language processing capabilities to handle nuanced concepts more effectively.

\section{Conclusion}

Our findings extend the theoretical understanding of AI bias beyond traditional demographic or political dimensions to encompass sustainability worldviews. The identification of distinct "sustainability personalities" among LLMs -- ranging from techno-pessimistic to techno-optimistic orientations --suggests that training data composition, model architecture, fine-tuning or imposed constraints shape not merely factual knowledge but fundamental perspectives on complex socio-technical challenges. This discovery contributes to the emerging literature on AI system evaluation and highlights the multidimensional nature of bias in LLMs.

The systematic differences we observed validate concerns about multi-stakeholder approaches to AI development that address embedded biases. LLMs embody distinct worldviews that influence their sustainability outputs, with profound implications for how organizations deploy these systems in contexts where sustainability considerations are paramount.

%The systematic differences we observed validate concerns about the need for multi-stakeholder approaches to AI development that acknowledge and address these embedded biases. These findings suggest that LLMs do not simply process information neutrally but instead embody particular worldviews that can significantly influence their outputs on sustainability-related topics. This has profound implications for how we conceptualize and deploy these systems in organizational contexts where sustainability considerations are paramount.

As organizations increasingly integrate LLMs into sustainability-related decision-making processes, our research underscores the critical importance of understanding each model's inherent perspectives. The polarized nature of LLM responses -- contrasting with humans' more nuanced views -- suggests that relying on any single model could lead to skewed strategic directions. Future research should explore the origins of these differences in perspectives, develop methods to mitigate their impact, and investigate how ensemble approaches might leverage diverse model viewpoints to support more balanced and effective sustainability initiatives. Ultimately, achieving the twin transition will require not just technological advancement but also careful consideration of the values and assumptions embedded within our AI tools.

\section*{Limitations}

This study has several limitations that should be considered when interpreting our findings. 

(1) The psychometric instruments employed were originally designed for human subjects, raising questions about their appropriateness for measuring AI biases. While the models produced consistent responses, some subscales showed low reliability, suggesting that traditional psychometric concepts may not translate directly to artificial systems. This is why the end-to-end use case simulation is of importance, which demonstrated that found biases translate into a practical application scenario. Further research should build on this and qualitatively evaluate further practical use cases (e.g., \citet{Giudici.2023}). 

Additionally, psychometric inventories themselves have inherent limitations when applied to understanding 'real' perspectives versus memorized survey responses. While these validated instruments provide standardized measurements, they may primarily capture surface-level patterns rather than deeper conceptual understanding. Our end-to-end budget allocation task partially addresses this concern by demonstrating that Likert-scale patterns translate to behavioral outcomes, but future research should explore whether these systematic differences reflect genuine conceptual frameworks or statistical artifacts from training data patterns.

(2) Our findings represent a temporal snapshot of current model capabilities. Given the rapid evolution of LLM architectures and training methodologies, these results may quickly become outdated as models are updated or retrained, potentially altering their sustainability perspectives.

(3) Our monolingual approach using English-language questionnaires may have introduced linguistic biases that affect how models interpret and respond to sustainability concepts. This constraint potentially limits our understanding of how these models would conceptualize sustainability when deployed in diverse linguistic and cultural contexts globally.

(4) While we maintained consistency by using the same set of prompts across all models, this standardized approach may have overlooked important architectural and operational differences between the systems. Future studies could improve accuracy by tailoring instructions to the specific design characteristics of each model, potentially revealing more nuanced variations in their sustainability conceptualizations.

(5) Our reliance on closed-ended Likert-scale responses represents a significant methodological constraint. This format restricts models to predetermined response categories and may not capture the full complexity of their sustainability conceptualizations. More realistic free-text responses might reveal different patterns or provide richer insights that contradict the systematic differences observed in our structured format. While this constraint was necessary for statistical comparability and to control API costs, it limits the ecological validity of our findings. Further, chain-of-thought prompting represents a promising complementary approach. Such prompting could provide richer insights into the reasoning processes underlying models' Likert-scale responses. Future research should systematically explore this method to enhance interpretability.

(6) Regarding our human baseline data (N=105), these participants were recruited through convenience sampling through social media, potentially limiting generalizability to broader populations. The human data was collected using identical instruments to ensure direct comparability with LLM responses, but demographic diversity and sample size constraints may affect the representativeness of our human-AI comparisons.

(7) We did not conduct mechanistic interpretability analysis to investigate the underlying computational causes of these differences, such as attention patterns or internal representations. Future work should incorporate such techniques to trace the origins of these sustainability biases, particularly for open-source models where internal components are accessible for analysis.

Despite these limitations, our findings provide valuable initial insights into how contemporary LLMs conceptualize sustainability.

\section*{Disclaimer: Use of assistive AI tools}
Generative AI tools were used to suggest non-substantive R code edits. The authors reviewed and verified all R code and outputs. No data were shared with the tool. Generative AI tools were also used for translation and language editing to enhance readability. The authors subsequently reviewed and revised the output as necessary and take full responsibility for the final content.

%\bibliographystyle{acl_natbib}
\bibliography{ACL_main}

\appendix
\renewcommand{\thetable}{\thesection\arabic{table}}

\section{Additional Results}\vspace{-10em}
\setcounter{table}{0}
\label{sec:appendixA}
\begin{table}[H]
\centering
\small
\caption{Mean and standard deviation of the models across all SDG17 criteria.}
\renewcommand{\arraystretch}{1.2}
\begin{tabular}{|l|c|c|}
\hline
\textbf{Model} & \textbf{Mean} & \textbf{SD} \\
\hline
GPT & 4.77 & 0.10 \\
CLAUDE & 4.70 & 0.06 \\
DEEPSEEK & 4.66 & 0.39 \\
LLAMA & 5.01 & 0.06 \\
MISTRAL & 4.16 & 0.10 \\
\hline
\end{tabular}
\label{tab:sdg17_performance}
\end{table}
\vspace{-4em}Contents of the SDG17 criteria as shown on the x-axis of Figure \ref{fig1}:\\\\
1: No Poverty\\
2: Zero hunger\\
3: Good health and well-being\\
4: Quality Education\\
5: Gender equality\\
6: Clean water and sanitation\\
7: Affordable and clean energy\\
8: Decent work and economic growth\\
9: Industry, innovation and infrastructure\\
10: Reduced inequalities\\
11: Sustainable cities and economies\\
12: Responsible consumption and production\\
13: Climate action\\
14: Life below water\\
15: Life on land\\
16: Peace, justice and strong institutions\\
17: Partnership for the goals

\section{Prompt Development}
\label{sec:appendixB}

\setcounter{table}{0}
\begin{table}[H]
\centering
\small
\caption{Prompt development and refinement process.}
\renewcommand{\arraystretch}{1.4}
\begin{tabular}{|p{1.7cm}|p{5cm}|}
\hline
\textbf{Version} & \textbf{Description} \\
\hline
\centering{V1} \newline Numerical Only & 
\textit{Prompt:} ``For each statement, indicate how well it describes you or your opinions. 
Select one of the following options: 1 = Strongly Disagree, …, 6 = Strongly Agree. Respond with the number only.'' \vspace{0.5em}\newline
\textit{Outcome:} Models often added explanations (``4 – Somewhat Agree because …'') or produced invalid/out-of-range numbers. \\
\hline
\centering{V2} \newline Categorical Labels & 
\textit{Prompt:} ``For each statement, indicate how well it describes you or your opinions. 
Select one of the following options: Strongly Disagree, Disagree, …, Strongly Agree.'' \vspace{0.5em}\newline
\textit{Outcome:} Cleaner outputs, but models frequently added explanations or disclaimers (``As an AI, I cannot …''). \\
\hline
\centering{V3} \newline Categorical\\\(+\) \\Basic Restrictions & 
\textit{Prompt:} ``For the following statement, you must respond with only one of the 
following options: Strongly Disagree … Strongly Agree. Do not elaborate, do not apologize, and do not add any other text.'' \vspace{0.5em}\newline
\textit{Outcome:} Reduced disclaimers, but some hedging language remained (``I somewhat agree, however …'') and occasional negative statements (``I cannot answer this …''). \\
\hline
\centering{V4} \newline Final Prompt & 
\textit{Prompt:} ``For each statement, indicate how well it describes you or your opinions. 
Select one of the following options: Strongly Disagree … Strongly Agree.'' Constraints: (1) No elaboration, (2) No alternative wording, (3) No apologies, (4) No disclaimers, (5) Exclude words like \textit{cannot, unable, instead, however, unfortunately}, (6) No negative sentences. \vspace{0.5em}\newline 
\textit{Outcome:} Produced highly consistent categorical responses, eliminated negative prompting (``I cannot…''), and ensured comparability. Numeric mapping was done only in post-processing. \\
\hline
\end{tabular}
\end{table}
\end{document}